\documentclass{ws-ijmpe}

\def\vec#1{\mbox{\boldmath $#1$}}
\newcommand{\vecr}{\vec{r}}

\begin{document}

\markboth{H. Oba and M. Matsuo}{Continuum Coupling and Pair Correlation in
Weakly Bound Deformed Nuclei}

\catchline{}{}{}{}{}

\title{CONTINUUM COUPLING AND PAIR CORRELATION IN WEAKLY BOUND DEFORMED NUCLEI}

\author{\footnotesize HIROSHI OBA}

\address{Graduate School of Science and Technology, Niigata University,
Niigata 950-2181, Japan\\
ooba@nt.sc.niigata-u.ac.jp}

\author{MASAYUKI MATSUO}

\address{Department of Physics, Faculty of Science,
Niigata University,
Niigata 950-2181, Japan\\
matsuo@nt.sc.niigata-u.ac.jp}

\maketitle


\begin{abstract}
We formulate a new Hartree-Fock-Bogoliubov method applicable to
weakly bound deformed nuclei using the coordinate-space 
Green's function technique.
An emphasis is put on treatment of quasiparticle states 
in the continuum, on which we impose the correct 
boundary condition of the asymptotic out-going wave. We illustrate
this method with numerical examples.
\end{abstract}

\section{Introduction}

The RI-beam facilities in the new generation
 will increase significantly the number of
experimentally accessible nuclei, especially  in medium and heavy
mass regions. We may reach nuclei close to the
neutron drip-line in the $10\le Z \le 20$ and $N\ge 20$ region, where
a bunch of deformed neutron-rich nuclei are expected\cite{Stoitsov}. 
This will provide us with a new opportunity to study
interplay among the 
presence of weakly bound neutrons, the coupling to
nuclear deformation effects, the pairing correlation, and the collective
excitations. 

A promising theoretical framework to describe this situation may be 
the self-consistent mean-field approaches.\cite{Bender}
More specifically, we consider here the Hartree-Fock-Bogoliubov 
(HFB) method to construct
the pair-correlated and deformed ground state, and 
the quasiparticle random phase approximation (QRPA) to
describe the excitation modes built on the ground state. 
As commonly recognized, one has to describe precisely 
the nucleon wave function of weakly bound and unbound orbits,
which should have proper asymptotic
behaviours. There exist such formulations for spherical 
nuclei,\cite{Belyaev,Dobaczewski,Grasso,Matsuo,Khan} but
a new challenge here is that we have to do it for deformed nuclei.
A method using the P\"{o}schel-Teller-Ginocchio basis is 
proposed recently\cite{Michel}. The quasiparticle motion
in deformed Woods-Saxon potential is analyzed in detail
in the coupled-channel formalism.\cite{Hamamoto}
We here take
a slightly different approach based on the 
coordinate-space Green's function
technique since we plan to apply it 
also to the continuum QRPA\cite{Matsuo}. 
In the present work, we shall show that the coordinate-space
Green's function technique enables us to formulate 
the deformed continuum HFB method in which the nucleon waves
satisfy a proper boundary condition of the asymptotic 
out-going wave. 

\section{Deformed continuum HFB method using the Green's function}

We first describe the quasi-particle motion in the HFB mean-fields consisting
of the particle-hole field and the pair field which are both deformed.
The axial symmetry is assumed. The quasiparticles
of the Bogoliubov type have two-component wave functions 
$\psi^{(1,2)}(\vecr\sigma)$, for which we use the radial coordinate system and the
partial wave expansion 
\begin{equation}
\psi^{(i)}(\vecr\sigma) = \sum_{L}^{L_{max}} \phi^{(i)}_L(r)y_L(\hat{\vecr}\sigma)
\end{equation}
with $L\equiv(jlm)$ and $y_L(\hat{\vecr}\sigma)$ being the spin spherical harmonics.
The HFB equation is then written as  
a coupled-channel Schr\"{o}dinger equations\cite{Hamamoto} for the 
radial wave functions $\{ \phi^{(i)}_L(r) \}$ 
where the quantum number $L$ represents
the ``channel''.  
Note that the energy spectrum of the quasiparticle
consists of  discrete and continuum parts, which are separated by
the energy condition $E<|\lambda|$ and $E>|\lambda|$ 
($\lambda$ is the Fermi energy) as is in the spherical case.\cite{Dobaczewski}

We can construct the exact Green's function for the
quasiparticle motion in our deformed HFB problem. It is an extension
of the spherical theory of Ref.\cite{Belyaev} to deformed cases, and we
accomplished 
this by employing a general prescription\cite{Foulis} of constructing the
exact Green's function for a deformed potential scatterer. Here the HFB Green's
function (a $2\times 2$ matrix form combining the normal and abnormal functions)
is expanded as
\begin{equation}
G(\vecr\sigma,\vecr'\sigma',E) = \sum_{L,L'}^{N_c} y_L(\hat{\vecr}\sigma)g_{LL'}(r,r',E)
y_{L'}^\dagger(\hat{\vecr}'\sigma').
\end{equation}
The coupled-channel radial Green's function
$ g_{LL'}(r,r',E)$ is constructed as a linear combination of 
products of ``regular solutions'' $\{ \phi^{{\rm I} (i)}_{LL'}(r) \}$
(i.e. those satisfying the boundary conditions 
$\phi^{{\rm I} (i)}_{LL'}(r) \rightarrow r^l \delta_{LL'}\delta_{ij}$
at the origin $r \rightarrow 0$) and 
 ``out-going wave solutions''   $\{ \phi^{{\rm O} (i)}_{LL'}(r) \}$
(those connected to the proper asymptotic form
$\phi^{{\rm O} (i)}_{LL'}(r) \rightarrow 
r^{-1}H_{l'}^+(k_ir) \delta_{LL'}\delta_{ij}$
for $r \rightarrow \infty$ where $H_{l'}^+(kr)$ is the out-going
Hankel function). 

We calculate the density
$\rho(\vecr)$ 
and the pair density $\tilde{\rho}(\vecr)$ using the HFB Green's function thus
constructed. The generalized density matrix 
\begin{equation}
R(\vecr\sigma,\vecr'\sigma') = 
\left(
\begin{array}{cc}
\rho(\vecr\sigma,\vecr'\sigma') & \tilde{\rho}(\vecr\sigma,\vecr'\sigma') \\
\tilde{\rho}^*(\vecr\tilde{\sigma},\vecr'\tilde{\sigma}')  & 
\delta_{\vecr\vecr'}\delta_{\sigma\sigma'}-\rho(\vecr\tilde{\sigma},\vecr'\tilde{\sigma}') 
\end{array}
\right)
= \frac{1}{2\pi i}\int_C G(\vecr\sigma,\vecr'\sigma',E) dE,
\end{equation}
which is a sum of the wave functions of all the quasiparticle
states including the continuum states,
is calculated using a contour integral of the
HFB Green's function\cite{Belyaev}. Incorporating this
way of calculating densities into the standard iterative algorithm,
we obtain the HFB ground state after convergence.
 
\begin{figure}
\psfig{file=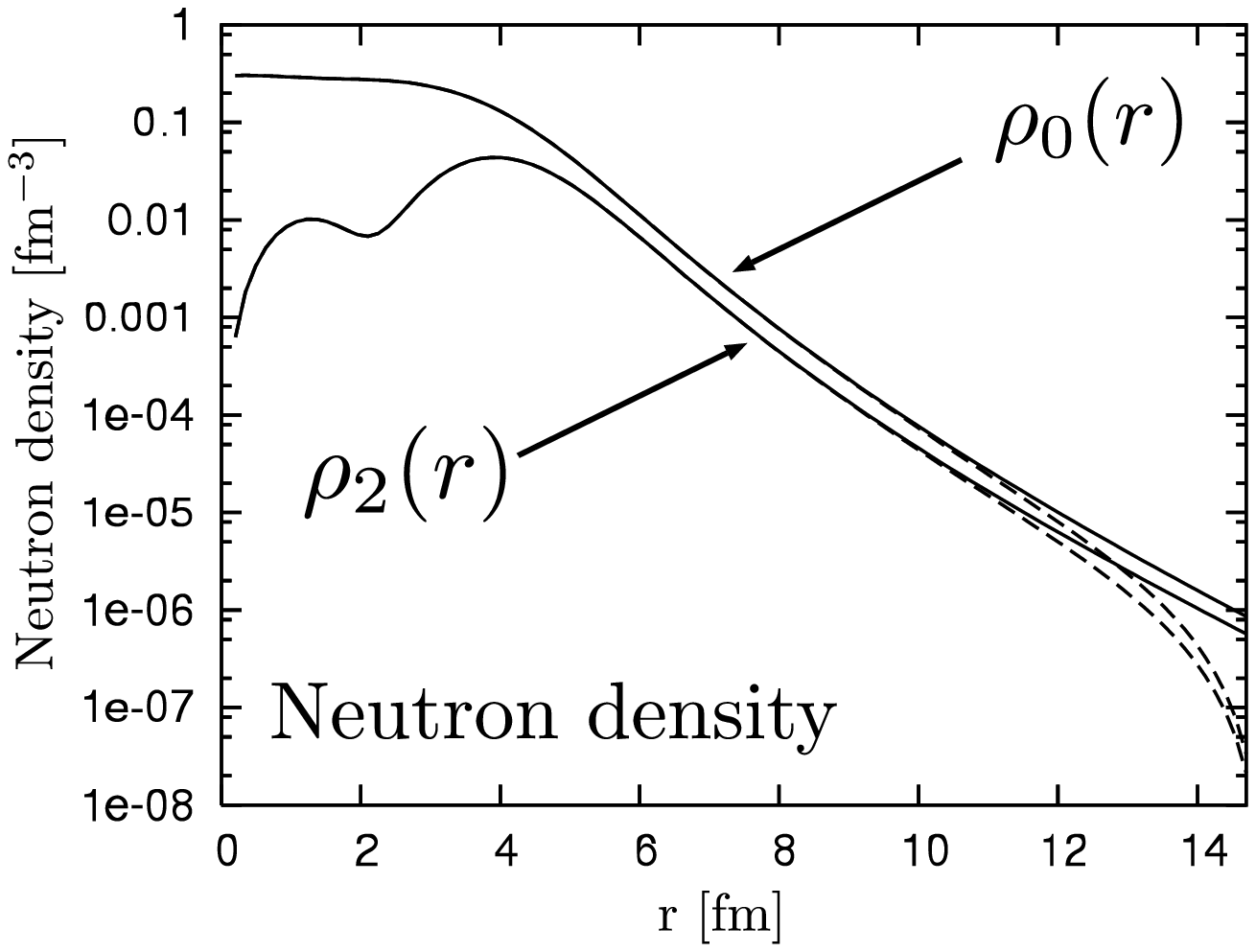,width=6cm}
\hspace{5pt}\psfig{file=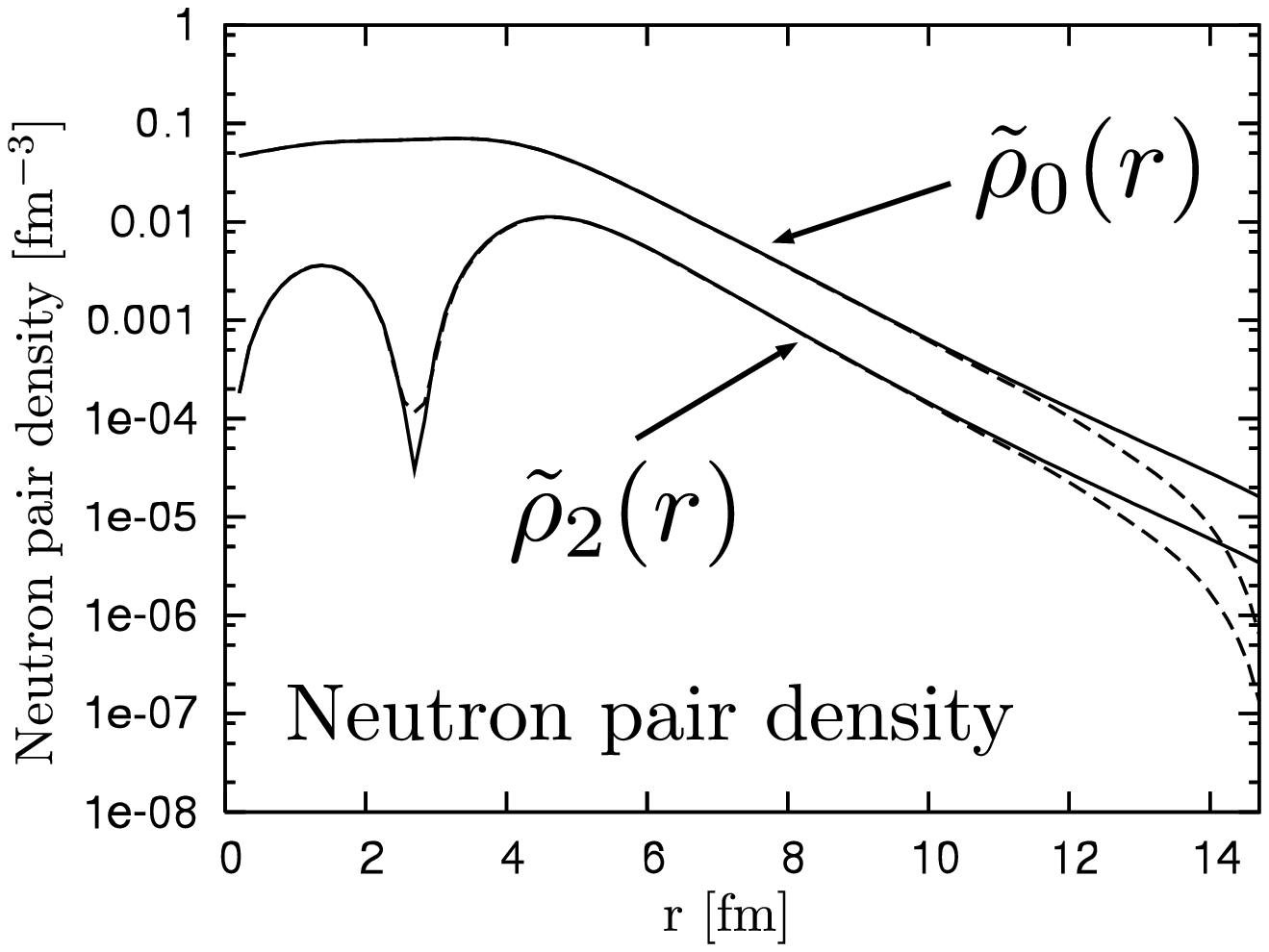,width=6cm}
\vspace*{9pt}
\caption{The monopole and quadrupole parts,  $\rho_0(r)$ and $\rho_2(r)$,
 of the neutron density (left panel). The monopole and
quadrupole parts,  $\tilde{\rho}_0(r)$ and $\tilde{\rho}_2(r)$,
of the neutron pair density (right). The dashed curves are the
results obtained with the box boundary condition.}
\end{figure}

\section{Numerical analysis}

We shall demonstrate with numerical examples how the deformed 
continuum HFB works. We adopt 
for simplicity a deformed Woods-Saxon potential as the 
particle-hole field, but we perform the HFB iteration
to obtain the selfconsistent pair field.  We use the density-dependent
delta interaction (DDDI) acting in the singlet pair, 
$
v_{pair}=v_0 \left(1-\eta(\rho_n(r)/0.08)^{0.59} \right) \delta(\vecr-\vecr'),
$
(for neutrons) 
where $v_0$ is fixed to reproduce the $nn$-scattering length $a=-18$ fm.
We consider $^{38}$Mg and assume a deformation $\beta=0.3$.  
Using the
Runge-Kutta-Nystrom method we solve numerically the coupled-channel equation  
within an interval $r=[0, r_{max}]$ ($r_{max}=15$  fm) with a step size
$\Delta r =0.2$ fm. 
At the outer boundary the wave functions are
connected to the asymptotic forms. 
The parameter $\eta=0.76$ is chosen to produce
the neutron pairing gap around $\Delta \sim 1.5 $ MeV.
The cut-off in the quasiparticle energy is
60 MeV, and the maximum $\Omega$ ($j_z$) quantum number is $\Omega_{max}=21/2$.
For comparison, we performed also the HFB calculation using the same model
but with a box boundary condition assuming an infinite wall at $r=r_{max}$. 
In the following we show results for neutrons.

Figure 1 shows the radial profile of the monopole and quadrupole parts 
$\rho_0(r)$ and $\rho_2(r)$ of the neutron density 
$\rho(\vecr)=\sum_\lambda\rho_\lambda(r)Y_{\lambda 0}(\hat{\vecr})$, 
and the corresponding
$\tilde{\rho}_0(r)$ and $\tilde{\rho}_2(r)$ of the neutron pair density
$\tilde{\rho}(\vecr)$. We obtain exponential asymptotics 
here thanks to the proper boundary condition, and it is 
in contrast to the results obtained with the box boundary condition
(the dashed curves in Fig.1). $\rho_0(r)$ and $\rho_2(r)$
have the same exponential slope, indicating that we can define the deformation
of the equi-density surfaces in the asymptotic region. 
This kind of deformed exponential tail 
is also seen in the neutron pair density. But the
ratio of $\tilde{\rho}_2(r)$ against $\tilde{\rho}_0(r)$ is significantly
smaller than that of the normal density. This points to that the
pair density in the tail has smaller deformation than that of the
normal density.

The quasiparticle spectrum above the threshold energy $E_{th}=|\lambda|$ 
should be 
continuous, and it is indeed the case in our formulation. 
Figure 2 shows the occupation number density $n(E)$ and the pair number density 
$\tilde{n}(E)$ which are defined by
\begin{equation}
   n(E) = \frac{1}{\pi}{\rm Im} \sum_{\sigma}\int d\vecr G_{11}(\vecr\sigma,\vecr\sigma,-E-i\epsilon), \ \ \ 
   \tilde{n}(E) = \frac{1}{\pi}{\rm Im}\sum_{\sigma} \int d\vecr G_{12}(\vecr\sigma,\vecr\sigma,-E-i\epsilon). 
\end{equation}
They quantify contributions of the quasiparticle state at energy $E$ 
to the neutron number $\int \rho(\vecr) d\vecr=N$ and to $\int \tilde{\rho}(\vecr) d\vecr$.
Here is shown only for the neutron $\Omega=1/2$ states. 
The smoothing parameter is chosen to $\epsilon=25$ keV 
(a discrete state would have artificial FWHM=50 keV).

\begin{figure}
\psfig{file=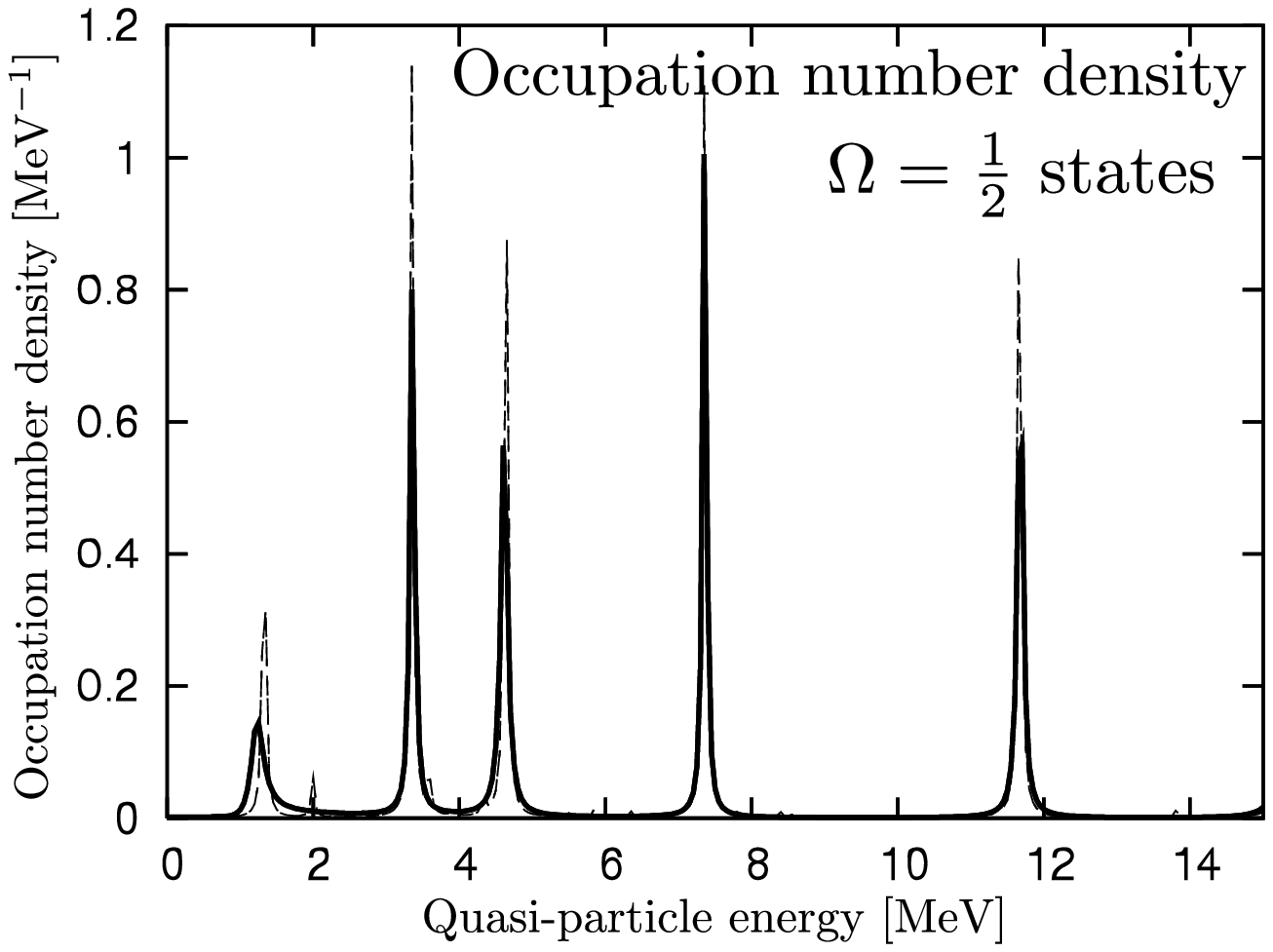,width=6cm}
\hspace{5pt}\psfig{file=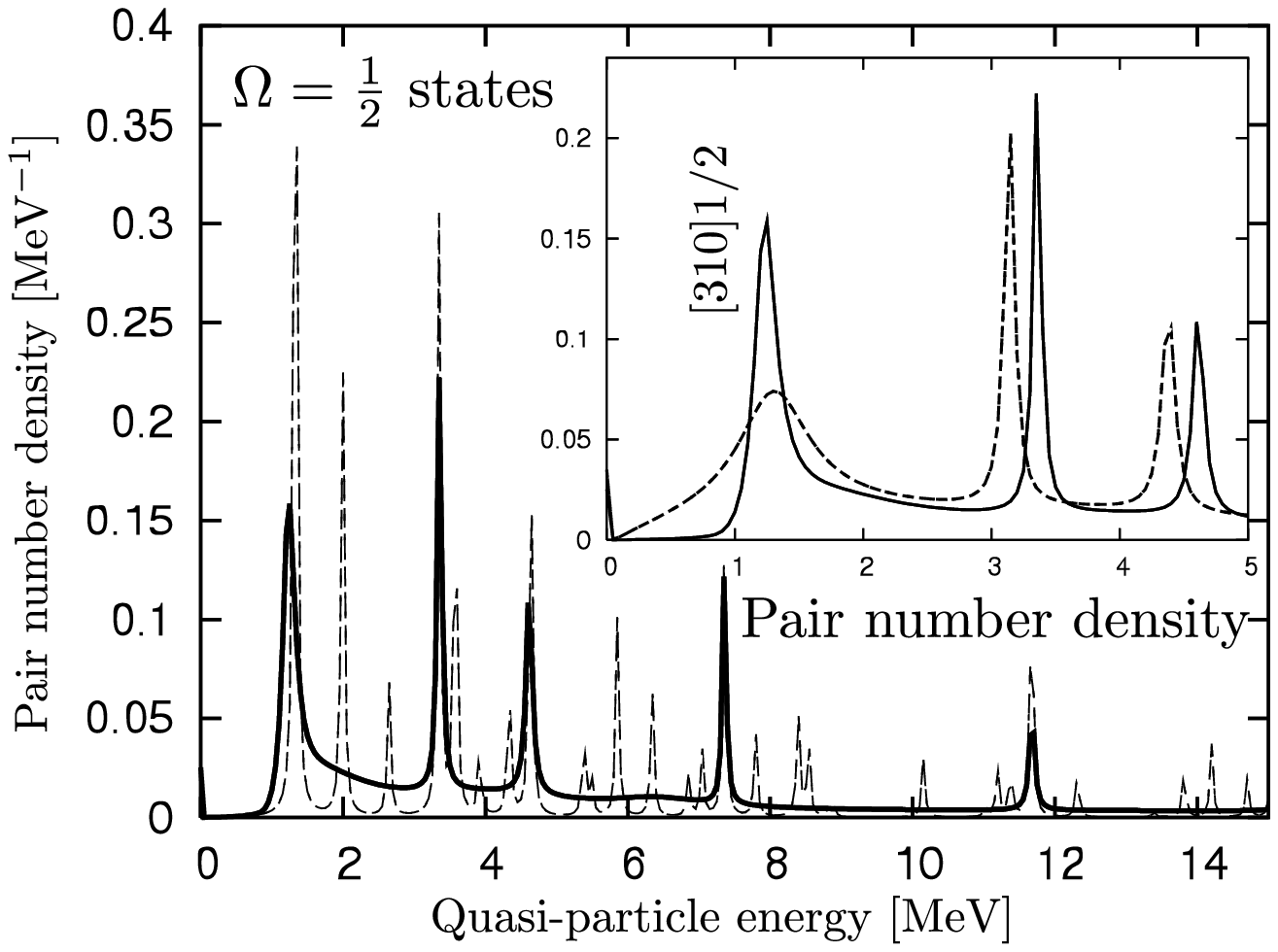,width=6cm}
\caption{The occupation number density $n(E)$ (left panel) and the pair
number density $\tilde{n}(E)$ (right panel) for neutrons, plotted with
the solid curves.
The results obtained with the box boundary condition are also plotted with the
dashed curves. The inset is a magnification of $\tilde{n}(E)$,
and we compare it with the result (the dotted curve) obtained with a
Woods-Saxon potential whose bottom is shifted up by +2 MeV.
  }
\end{figure}

In this numerical example
there is no
discrete quasiparticle states below $|\lambda|$ ($\lambda=-889$ keV),
and all the quasiparticle states are
embedded in the continuum $E> |\lambda|$. The quasiparticle 
states corresponding
to deep hole Woods-Saxon orbit appear 
both in $n(E)$  and $\tilde{n}(E)$ as narrow resonances.
It is observed also that the non-resonant
continuum states have a significant contribution 
to the pair number density  $\tilde{n}(E)$. 
This figure shows also that the box-discretized
calculation has difficulty to describe the non-resonant continuum states. 
The lowest energy resonance is not described well
by a single state in the box-discretized calculation, and it is because this
resonance has a rather large width. This
resonance corresponds to the $[310]\frac{1}{2}$ orbit 
in the deformed Woods-Saxon potential,
which is, in the absence of the pairing, a bound state with the single-particle
energy $e=-798$ keV. 

Naturally we expect most dramatic effect of the weak binding 
on this state. The inset shows how the spectrum $\tilde{n}(E)$ changes when
the depth of the Woods-Saxon potential is artificially shifted (made shallower)
by 2 MeV. The Woods-Saxon single-particle  energy $e$ and the Fermi energy
$\lambda$ changes from $e,\lambda=-798, -889$ keV 
to $e,\lambda=-40, -88$ keV.
We see in the inset of Fig.2 
a dramatic increase in the width of the lowest-energy resonance, which
apparently originates from the weak binding.
Note however that the peak energy of the resonance
stays almost constant. This implies that the effective
pairing gap of this resonant quasiparticle state is unchanged,
if we estimate the effective pairing gap
through the relation $E_{qp}=\sqrt{(e-\lambda)^2+\Delta^2}$.
This observation is
different from that in Ref.\cite{Hamamoto} claiming 
a reduction of the effective
pairing gap due to the weak binding effect. The difference originates from
the fact that we here use the selfconsistent pairing field 
generated from the DDDI,
whose force strength becomes large at low densities. 
The pair field extending to far outside the nucleus plays important roles.

\section{Conclusions}
We have formulated the deformed continuum HFB method 
which is designed for weakly bound deformed nuclei. We utilized here
the exact construction of the quasiparticle Green's function for
deformed HFB mean-fields on the basis of the coupled-channel representation.
The proper boundary condition of the out-going wave is imposed on the
continuum quasiparticles. We have analyzed numerically effects
of the continuum coupling and the weak binding on the pair correlation.
It is found that the quasiparticle states
in the non-resonant continuum play significant role to generate the pair
correlation. It is also suggested that the effective
pairing gap of weakly bound orbits is not reduced very much,
provided that the pairing interaction has the surface enhancement.

\section*{Acknowledgments}
The work is supported by
the Grant-in-Aid for Scientific Research(No.20540259) from the Japan
 Society for the Promotion of Science,  and also by 
the JSPS Core-to-Core Program, International
Research Network for Exotic Femto Systems(EFES).


\begin{thebibliography}{0}
\bibitem{Stoitsov} M. V. Stoitsov, J. Dobaczewski, W. Nazarewicz, S. Pittel
and D. J. Dean, {\it Phys. Rev. C} {\bf 68} (2003) 054312.

\bibitem{Bender}
M.~Bender, P.~-H.Heenen, P.~-G.~Reinhard, {\it Rev. Mod. Phys.} {\bf 75} (2003) 121.

\bibitem{Belyaev} 
S.~T.~Belyaev, A.~V.~Smirnov, S.~V.~Tolokonnikov, S.~A.~Fayans,
{\it Sov. J. Nucl. Phys.} {\bf 45} (1987) 783.


\bibitem{Dobaczewski} 
J.~Dobaczewski, H.~Flocard, and J.~Treiner,
{\it Nucl. Phys. A} {\bf 422} (1984) 103.


\bibitem{Grasso} 
M.~Grasso, N.~Sandulescu, Nguen~Van~Giai, and R.~J.~Liotta,
{\it Phys. Rev. C} {\bf 64} (2001), 064321.

\bibitem{Matsuo} 
M.~Matsuo, {\it Nucl. Phys. A} {\bf 696} (2001) 371.

\bibitem{Khan}
E.~Khan, N.~Sandulescu, M.~Grasso, and Nguyen~Van~Giai,
{\it Phys. Rev. C} {\bf 66} (2002) 024309.

\bibitem{Michel}
M.~Stoitsov, N.~Michel, and K.~Matsuyanagi,
{\it Phys. Rev. C} {\bf 77} (2008) 054301.

\bibitem{Hamamoto}
I.~Hamamoto, {\it Phys. Rev. C} {\bf 71} (2005) 037302; {\bf 73} (2006) 044317.


\bibitem{Foulis}
D.~L.~Foulis, {\it Phys. Rev. A} {\bf 70} (2004) 022706.


\end{thebibliography}
\end{document}